\documentclass[prb,aps,twocolumn]{revtex4}
\usepackage{bm,graphicx,amsmath} \usepackage{bbm}
\newcommand{\abs}[1]{\left|{#1}\right|}

\newcommand{\br}[1]{\langle #1|}
\newcommand{\ke}[1]{|#1\rangle}

\newcommand{\kb}[2]{\ke{#1}\br{#2}}

\newcommand{\al}[1]{^{(#1)}}
\newcommand{\da}{^\dagger}

\newcommand{\pt}[1]{\left( #1 \right)}
\newcommand{\pq}[1]{\left[ #1 \right]}

\begin{document} \title{Ground-state cooling the vibrations of suspended carbon-nanotubes with constant electron current} \author{Stefano Zippilli,$^{1,2,3}$ Adrian Bachtold,$^4$ and Giovanna Morigi$^{1,2}$} \affiliation{ $^{1}$Departament de F\'{i}sica,
Universitat
Aut\`{o}noma de Barcelona, E-08193 Bellaterra, Spain,\\
$^2$ Theoretische Physik, Universit\"at des Saarlandes, 66041 Saarbr\"ucken, Germany,\\
$^3$ Fachbereich Physik, Universit\"at Kaiserslautern, 67663 Kaiserslautern, Germany,\\
$^4$ CIN2(CSIC-ICN), Campus de la UAB, E-08193 Bellaterra, Spain.} \date{\today}

\begin{abstract}
We investigate the efficiency of cooling the vibrations of a nano-mechanical resonator, constituted by a partially suspended Carbon-nanotube and operating as double-quantum dot. The motion is brought to lower temperatures by tailoring the energy exchange via electromechanical coupling with single electrons, constantly flowing through the nanotube when a constant potential difference is applied at its extremes in the Coulomb-blockade regime. Ground-state cooling is possible at sufficiently high quality factors, provided that the dephasing rate of electron transport within the double dot does not exceed the resonator frequency. For large values of the dephasing rates cooling can still be achieved by appropriately setting the tunable parameters.
\end{abstract}

\maketitle

\section{Introduction}
The experimental demonstration of strong coupling between a nano-mechanical resonator and charge transport in carbon nanotubes~\cite{Lassagne,Steele} holds promising perspectives for cooling the temperature of the resonator vibrations by means of electro-mechanical forces. It may enable one to improve the sensitivity of high-precision measurements of mass~\cite{Roukes, Zettl, Benjamin, Bockrath}, mechanical displacement~\cite{Lahaye}, or spin~ \cite{Rugar}. Cooling by means of electro-mechanical coupling, moreover, may allow one to prepare the resonator state in the quantum regime~\cite{Schwab-Roukes}. Recently, it has been theoretically proposed~\cite{Martin,Blencowe,Clerk,MartinPZ04,Pistolesi,Nori,Ouyang,Sonne} and experimentally demonstrated~\cite{Naik} that back-action cooling can be achieved by coupling mechanical resonators to the constant electron current flowing through electronic nano-devices, such as normal-metal and superconducting single-electron transistors~\cite{LaHaye09}. Within these studies, it is found that one of the major factors limiting the cooling efficiency is the quality factor of the mechanical resonator. In this respect a suspended carbon nanotube is a promising candidate for achieving ground state cooling~\cite{Huttel09}.

\begin{figure}[!th]
\begin{center}
\includegraphics[width=8cm]{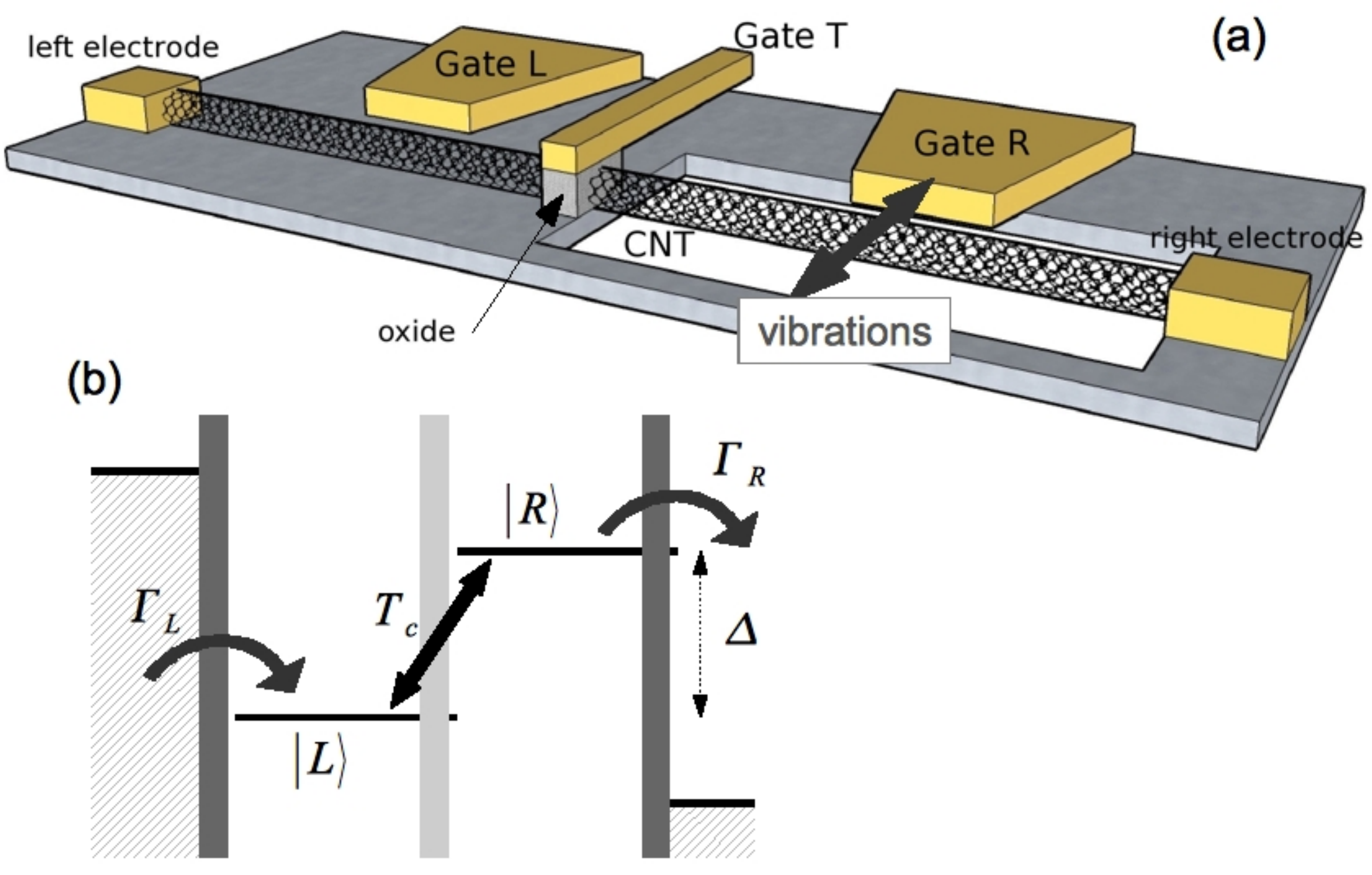}
\caption{(Color online) (a) Sketch of the considered set-up. The double quantum dot is a carbon nanotube (CNT) operating as single-electron transistor. In the picture, the dot on the right is suspended and mechanically vibrates. Its vibrations couple to the electronic current via electron-phonon interaction. In this article we will also consider the situation in which it is the dot on the left which is suspended. (b) Energy diagram for the DQD. States $\ke{L}$ and $\ke{R}$ denote one excess electron in the left and right dot, respectively. The two levels are at tunable energy difference $\hbar \Delta$, and are coupled by a tunneling barrier with tunneling matrix element $T_c$. Left (right) electrode act as source (drain) such that electron tunnels from left to right at rates $\Gamma_L$ and $\Gamma_R$. }
\label{figlevels}
\end{center}
\end{figure}

In this work we theoretically analyze the efficiency of ground-state cooling the vibrations of a nano-mechanical resonator using constant electron current, in a setup like the one sketched in Fig.~\ref{figlevels}. Here, the resonator is a suspended nanotube, which constitutes also the electronic device through which the current flows, in a setup which is reminescent to the ones realized in Refs.~\cite{Lassagne,Steele}. Moreover, the nanotube acts as a double quantum dot (DQD). This setup allows us to tune the electron current, so to enhance the electromechanical processes which irreversibly absorb phonons from the resonator. The analysis here presented extends and complements the proposal in Ref.~\cite{Zippilli09}, where we predicted ground-state cooling in this system for a specific set of parameters. In the present work we study the cooling efficiency over a larger parameter space than the one identified in Ref.~\cite{Zippilli09}, showing that large ground-state occupations can also be obtained for relatively large tunneling rates. Moreover, we analyze in detail the role of the various noise sources on the cooling efficiency.

It must be mentioned that cooling and manipulation of nano-mechanical resonators using electro-mechanical forces is complementary to the approach using the coupling to photons~\cite{Kippenberg,Zeilinger,Heidmann,Bouwmeester,Lehnert,Schliesser,Park,Groblacher,Rocheleau,Wilson-Rae09}. Both approaches are presently experimentally pursued. In particular, cooling by means of constant electron currents is appealing because it is easy to implement in a dilution fridge as compared to techniques based on photons. Beyond the specific implementation, ground-state cooling of the mechanical resonator would open the possibility to create and manipulate non classical states at the macroscopic scale and to study the transition from the classical to the quantum regime~\cite{Schwab-Roukes,SchwabEntanglement,Rabl04}.

This article is organized as follows. In Sec.~\ref{Sec:2} the model is introduced, which describes the coherent and incoherent dynamics of the nanotube. The rate equations for cooling are derived in Sec.~\ref{Sec:3}, and the cooling efficiency is discussed as a function of the physical parameters. The conclusions are presented in Sec.~\ref{Sec:4}.

\section{The model}
\label{Sec:2}

\subsection{Setup and basic dynamics}

The resonator we consider corresponds to a fundamental vibrational eigenmode of a partially suspended carbon nanotube (CNT). The eigenfrequency of the mode is $\omega$, and the resonator is assumed to be characterized by sufficiently high quality factors, so that the eigenmode can be selectively addressed by suitably tailoring the mechanical coupling with the constant electronic current flowing though the CNT. The eigenmode is here described by a quantum harmonic oscillator, with $a\da$ and $a$ the bosonic creation and annihilation operators of a quantum of energy $\hbar\omega$. The Hamiltonian for the resonator is given by
$$H_{ph}=\hbar \omega a\da a\,,$$
and it governs its coherent dynamics. The eigenstates are the number states $|n\rangle$, with energy $\hbar n\omega$ and $n=0,1,2,\ldots$

Before discussing the details of the electromechanical coupling, we first introduce the electronic properties of the system. The system is in a double quantum dot (DQD) configuration, as sketched in Fig~\ref{figlevels}(a): Voltage applied to gate T provides the potential barrier, which divides the CNT in two coupled quantum dots~\cite{Marcus,Kouwenhoven,Schonenberger,Lindelof}. The voltage applied at the left and right electrodes warrants that a constant electron current flows through the CNT from left to right. We denote by  $\mu_L$ and $\mu_R$ the chemical potential at the left and right electrode, respectively, such that $\mu_L>\mu_R$. We assume that the DQD is working in the Coulomb-blockade regime such that the charging energy $E_c$ is much larger then the transport energy window, $E_c\gg\mu_L-\mu_R$. In this regime the system operates as a single-electron transistor (SET), and the relevant electronic states are the ground state of the DQD, $\ke{0}$, and one excess electron either in the left or in the right dot, corresponding to the quantum states $\ke{L}$ and $\ke{R}$, respectively. More specifically, the electronic ground state corresponds to the state with $N_L$ and $N_R$ electrons in the left and right QD respectively, $\ke{0}\equiv\ke{N_L,N_R}$, while $\ke{L}\equiv\ke{N_L+1,N_R}$ and $\ke{R}\equiv\ke{N_L,N_R+1}$, see for instance Refs.~\cite{Brandes05,Nazarov,KouwenhovenReview}. We denote by $\epsilon_L$ and $\epsilon_R$ the energy of the states $\ke{L}$ and $\ke{R}$ with respect to the ground state. Their difference is controlled by the gates L and R in Fig~\ref{figlevels}(a). The DQD Hamiltonian, governing the dynamics of the electrons in the DQD, reads
\begin{eqnarray}\label{HDQD}
    H_{DQD}=H_{LR}+H_T\,,
\end{eqnarray}
with
\begin{eqnarray}
&&H_{LR}=\epsilon_L\kb{L}{L}+\epsilon_R\kb{R}{R}\,,\\
&&H_T=-\hbar T_c\pt{\kb{R}{L}+\kb{L}{R}}\,,
\end{eqnarray}
where $T_c$ is the tunneling matrix element between left and right QD and has the dimension of a frequency. The energy levels of the QDs and the direction of the electron current are depicted in  Fig~\ref{figlevels}(b).

We now discuss the electromechanical coupling. The vibration are capacitively coupled to the gate R: The gate voltage bends the CNT influencing its electric and mechanical properties. When the vibration amplitude is sufficiently small, the electromechanical energy can be considered to be linear in the amplitude. Assuming that the energy difference between left and right dot is such that $\epsilon_R-\epsilon_L\sim \hbar \omega$, the electronic transport couples with the fundamental eigenmode at frequency $\omega$, and this mode is displaced along $x$ so that the Hamiltonian describing this dynamics can be approximated by the operator
\begin{eqnarray}\label{He-ph}
    H_{e-ph}=\hbar \alpha\kb{R}{R}\pt{a\da+a}\,,
\end{eqnarray}
where we used $\hat x\propto a\da+a$, and $\alpha$ has the dimension of a frequency and gives the strength of the electromechanical coupling. This Hamiltonian describes a shift of the center of oscillation of the harmonic vibrations, conditioned to the occupation of the electronic state $\ke{R}$ of the DQD or, in other terms, an energy shift of the state $\ke{R}$ proportional to the mechanical displacement of the resonator.

We finally introduce the model for the coupling of the DQD with the electrodes, which gives rise to the constant electron current flowing through the CNT. The electrodes are electron reservoirs, described by the Hamiltonian
\begin{eqnarray}
H_{\rm res}=\sum_k \epsilon_k^L c_k\da c_k+\sum_k
\epsilon_k^R d_k\da d_k\,,
\end{eqnarray}
where $c_k$ and $d_k$ ($c_k\da$ and $d_k\da$) are fermionic operators, which annihilate (create) an electron of the respective reservoir and $k$ runs over all modes of the electrons. Electrodes and QDs are coupled by electron tunneling, which in Hamiltonian form reads
\begin{eqnarray}
W_L&=&\sum_k V_k^L c_k\da s_L+{\rm H.c.}\,,\\
W_R&=&\sum_k V_k^R d_k\da s_L+{\rm H.c.}\,, \label{WR}
\end{eqnarray}
where $V_k^L$ and $V_k^R$ are the tunneling matrix elements between reservoirs and quantum dots, and $s_\xi=\kb{0}{j}$ annihilates an electron in the QD $j$, with $j=L,R$, leaving the QD in the ground state while an electron is created in one mode of the reservoir.

In what follows we assume that the bias voltage is sufficiently large such that thermal effects in the coupling with the electrodes are negligible.
This implies that the relations $\mu_L-\mu_R\gg k_B T$ and $\mu_L\gg\epsilon_L,\epsilon_R\gg\mu_R$ must hold. In this regime the electrons tunnel from left to right, while transport in the opposite direction can be neglected. Tunneling from and into the electrodes can be described in perturbation theory. In first-order
perturbation theory the tunneling rate $\Gamma_L$ from the left electrode into the DQD, and the tunneling rate $\Gamma_R$ from the DQD to the right electrode, are given by
\begin{eqnarray}
&&\Gamma_{L}=\frac{2\pi}{\hbar}\sum_k \abs{V_k^{L}}^2\delta\pt{\epsilon_{L}-\epsilon_k^{L}}\,,\\
&&\Gamma_{R}=\frac{2\pi}{\hbar}\sum_k \abs{V_k^{R}}^2\delta\pt{\epsilon_{R}-\epsilon_k^{R}}\,,
\end{eqnarray}
where $\delta(\epsilon)$ expresses the condition of energy conservation between initial and final state. In the continuum limit the sums over the modes of the electrodes become integrals, properly weighted by the density of states.

\subsection{Master equation}

In order to describe the effect of quantum noise on the resonator dynamics we resort to the density matrix for the degrees of freedom of the mechanical resonator and of the DQD electronic degrees of freedom. Its dynamics is governed by the master equation, which also accounts for incoherent tunneling from and into the electrodes, dephasing of trasport between the QDs, and thermalization of the resonator to the temperature at the electrodes. The master equation reads
\begin{eqnarray}\label{Meq}
    \frac{\partial}{\partial t}\rho&=&\frac{1}{{\rm i} \hbar}\pq{H_{DQD}+H_{ph}+H_{e-ph},\rho}\\
& &+{\cal L}_{\rm DQD}\rho
+{\cal L}_d\rho+{\cal K}\rho\,,\nonumber
\end{eqnarray}
where the term
\begin{eqnarray}
    {\cal L}_{\rm DQD}\rho&=&\Gamma_L/2\pt{2s_L\da\rho s_L-\rho s_Ls_L\da-s_Ls_L\da\rho}
    \nonumber   \\&&
    + \Gamma_R/2\pt{2s_R\rho s_R\da-\rho s_R\da s_R-s_R\da s_R\rho}
\end{eqnarray}
describes incoherent electron tunneling from left electrode to state $\ke{L}$, and from state $\ke{R}$ to the right electrode, while the term
 \begin{eqnarray}
    {\cal L}_d=\frac{\Gamma_d}{4}\pt{2s\al{z}\rho s\al{z}-{s\al{z}}^2\rho-\rho{s\al{z}}^2}
\end{eqnarray}
describes electronic dephasing inside the DQD at rate $\Gamma_d$, with $s\al{z}=\kb{R}{R}-\kb{L}{L}$. Finally
\begin{eqnarray}\label{K}
    {\cal K}\rho&=&(\bar n_p+1)\gamma_p/2\pt{2a\rho a\da-a\da a\rho-\rho a\da a}\nonumber\\
    &&+\bar n_p\gamma_p/2 \pt{2a\da\rho a-aa\da\rho-\rho aa\da}
\end{eqnarray}
represents processes leading to thermalization at rate $\gamma_p$ with a thermal environment at temperature $T$, such that the mean number of phonon at frequency $\omega$ is given by $$\bar n_p=\pq{{\rm exp}\pt{\hbar\omega/k_B T}-1 }^{-1}\,.$$ The Liouvillian terms in Eq.~(\ref{Meq}) are written in the Born-Markov approximation: The master equation is local in time, it hence does not consider memory effects which may arise from the coupling with the reservoir. A master equation for a double-quantum dot coupled to a resonator has been reported in Refs.~\cite{Brandes05,Brandes}. With respect to that master equation, here we have added the finite temperature of the electrodes and the dephasing mechanism. The latter is here modeled according to the master equation presented for instance in Ref.~\cite{Viola}.

\section{Electro-mechanical cooling}
\label{Sec:3}

Ground-state cooling via electro-mechanical coupling can be achieved by enhancing the processes of electron transport which absorb a phonon of the nano-mechanical resonator. This is realized by properly setting the energy difference between the two QDs. Such processes must take place at rates which overcome the rate of electro-mechanical processes which tend to heat the resonator and the rate at which the resonator naturally thermalizes with the surrounding environment. In this section we derive the basic equations which describe the cooling dynamics of the resonator, and study them in different limits. The validity of the analytical predictions is tested by comparing them with the numerical solution of the master equation.

\subsection{Rate Equations for cooling dynamics}

In order to identify the important parameters, which determine the cooling dynamics, we resort to perturbation theory and derive rate equations for the occupation of the resonator's vibrational states. This is performed eliminating the electronic degrees of freedom from the resonator dynamics, assuming that $\alpha$, scaling the electromechanical coupling, is much smaller than the characteristic rates determining the electronic dynamics. In this regime, we assume that the electronic degrees of freedom of the DQD are in the steady state. This steady state determines the properties of the electronic current, and thus also the autocorrelation spectrum of the electro-mechanical force acting on the resonator. Assuming time-scale separation between the electronic dynamics and the dynamics of electromechanical coupling, we then derive a master equation for the resonator degrees of freedom, obtained from master equation~(\ref{Meq}) in second order perturbation theory and tracing out the electronic variables. For this purpose we rewrite the master equation~(\ref{Meq}) as
\begin{equation}\label{MEq:1}
\frac{\partial}{\partial t}\rho=\left({\mathcal L}_0+{\mathcal L}_1+{\mathcal K}\right)\rho\,,
\end{equation}
where
\begin{eqnarray}\label{L0}
\mathcal L_0\rho={\cal L}_0\al{0}\rho-\frac{{\rm i}}{\hbar}\pq{H_{ph},\rho}\,,
\end{eqnarray}
accounts for DQD and resonator dynamics in absence of electro-mechanical coupling, with
\begin{eqnarray}\label{L00}
{\cal L}_0\al{0}\rho=-\frac{\rm i}{\hbar}\pq{H_{DQD},\rho}+\pt{{\cal L}_{\rm DQD}+{\cal L}_{d}}\rho
\end{eqnarray}
the Liouvillian for the DQD. The term $\mathcal L_1$ contains the electromechanical coupling term,
\begin{eqnarray}
{\mathcal L}_1\rho=-\frac{\rm i}{\hbar}[H_{e-ph},\rho]\,,
\end{eqnarray}
and it scales with $\alpha$. Finally, the Liouvillian $\cal K$ describes the thermalization of the resonator, it scales with $\gamma_p$, and is given in Eq.~(\ref{K}).

We assume that the back-action of the electromechanical coupling on the electronic state can be neglected, which corresponds to considering the regime where $|\alpha|\ll\Gamma_R,\Gamma_L,|T_c|$. Then, the electronic steady state $\rho_{\rm St}\al{0}$ is determined by the solution of the equation
\begin{eqnarray}\label{StSt}
    {\cal L}_0\al{0}\rho_{\rm St}\al{0}=0\,.
\end{eqnarray}
In particular, $\rho_{\rm St}\al{0}$ can be written as
\begin{eqnarray}
\rho_{\rm St}\al{0}&=&\rho_{00}\ke{0}\br{0}+\rho_{LL}\ke{L}\br{L}+\rho_{RR}\ke{R}\br{R}\nonumber\\
&&+\rho_{RL}\ke{R}\br{L}+\rho_{LR}\ke{L}\br{R}
\end{eqnarray}
where the elements scaling with the parameters $\rho_{0R}$ and $\rho_{0L}$, and corresponding to the coherences between vacuum and $\ke{L}$, $\ke{R}$, are not reported, as $\rho_{0L}$ and $\rho_{0R}$ vanish at steady state due to incoherent tunneling between the electrodes and the DQD. The other parameters take the form
\begin{eqnarray}\label{rhoStSt:1}
\rho_{LL}&=&\frac{1}{\frac{\Gamma^2}{4}+\Delta^2}\pq{2\frac{\Gamma_d' T_c^2}{\Gamma_R}+{\Gamma_d'}^2+\Delta^2}\,,\\
\rho_{RR}&=&2\frac{\Gamma_d'T_c^2}{\Gamma_R \pt{\frac{\Gamma^2}{4}+\Delta^2}}\,,\\
\rho_{RL}&=&\frac{T_c}{\frac{\Gamma^2}{4}+\Delta^2}\pt{\Delta+{\rm i}\Gamma_d'}\,,
\label{rhoStSt:3}
\end{eqnarray}
with $\rho_{LR}=\rho_{RL}^*$ and $\rho_{00}=1-\rho_{RR}-\rho_{LL}$. In Eqs.~(\ref{rhoStSt:1})-(\ref{rhoStSt:3}) we defined $\Gamma=2\pq{\Gamma_d'\pt{\frac{4T_c^2}{\Gamma_R}+\frac{2T_c^2}{\Gamma_L}+\Gamma_d'}}^{1/2}$  and
\begin{equation}\label{Gamma:d}
\Gamma_d'=\Gamma_d+\Gamma_R/2\,,
\end{equation}
giving the total decay of the electronic coherences of the DQD. We further assume that $$|\alpha|\ll\omega\,.$$ In this limit, a closed set of equations for the occupation of the resonator states can be derived. For this purpose, we define the projector $P$ acting over the density matrix of the system, such that its action over a given density matrix $\rho$ of the system reads
\begin{eqnarray}\label{P}
P\rho=\rho_{\rm St}\al{0}\otimes{\rm Tr}_{\rm el}\{P_0\rho\}\,,
\end{eqnarray}
where ${\rm Tr}_{\rm el}$ denotes the trace over the electronic degrees of freedom and $P_0\rho=\sum_n|n\rangle\langle n|\langle n|\rho|n\rangle$.
The master equation for the density operator projected over this subspace can be written in the form~\cite{Cirac92,Morigi03}
\begin{eqnarray}\label{Pdotrho}
\frac{\partial}{\partial t}P\rho&=&P{\mathcal K}P\rho+P\int_0^{\infty}{\rm d}\tau{\mathcal L}_1{\rm e}^{{\cal L}_0\al{0}\tau}{\mathcal L}_1P\rho(t)\,,
\end{eqnarray}
where we applied a perturbation expansion to second order in $\alpha$, and took $\gamma_p\simeq {\rm o}(\alpha^2)$, so that thermalization effects are taken at lowest order in the expansion. This latter assumption is indeed important for the efficiency of the cooling process, as only in this limit the electromechanical coupling may be able to counteract effectively thermalization with the external environment, bringing the resonator into a dynamical equilibrium with the electronic current.

After tracing out the electronic degrees of freedom from Eq.~(\ref{Pdotrho}) one obtains a set of rate equations for the occupation $p_n$ of the resonator state with $n$ phononic excitations, which read~\cite{Stenholm,Eschner}
\begin{eqnarray}\label{rateEq}
\dot{p}_n=(n+1)A_-p_{n+1}-\left[(n+1)A_++nA_-\right]p_n+nA_+p_{n-1}\,,\nonumber\\
\end{eqnarray}
where $A_{+}$ and $A_-$ are the rate for processes which increase and decrease, respectively, the state of the mechanical resonator by one excitation. They result from the interplay between the thermalization processes and of the cooling due to the mechanical effects induced by the electron current. Within the parameter regimes considered so far, they are the sum of thermalization and electro-mechanical rates, and written as
\begin{eqnarray}
A_{-}&=&\gamma_p(\bar n_p+1)+A_{0-}\,,\\
A_{+}&=&\gamma_p\bar n_p+A_{0+}\,,
\end{eqnarray}
where $A_{0+}=2{\rm Re}\{S(-\omega)\}$ ($A_{0-}=2{\rm Re}\{S(\omega)\}$) is the rate for electromechanical heating (cooling), with
\begin{eqnarray}\label{A0pm}
S(\omega)=-\alpha^2{\rm Tr}\left\{|R\rangle\langle R|\left({\cal L}_0\al{0}+{\rm i}\omega\right)^{-1}|R\rangle\langle R|\rho_{\rm St}\al{0}\right\}
\end{eqnarray}
the spectrum of the autocorrelation function of the electromechanical coupling force~\cite{Cirac92,Morigi03,Zippilli05}. It can be rewritten as
\begin{eqnarray}
S(\omega)&=&\rho_{RR}S_{RR}(\omega)+\rho_{RL}S_{RL}(\omega)\,,
\end{eqnarray}
where
\begin{eqnarray}
S_{RR}(\omega)
& =&\frac{\alpha^2}{2F(\omega)}\pq{\pt{\Gamma_d'-{\rm i}\omega}^2+2{\rm i}\frac{T_c^2}{\omega}\pt{\Gamma_d'-{\rm i}\omega} +\Delta^2     } \,,\nonumber\\
S_{RL}(\omega)
&=&-{\rm i}\frac{\alpha^2 T_c}{2F(\omega)}\pq{\Gamma_d'-{\rm i}\pt{\omega+\Delta}}\,,
\end{eqnarray}
and $F(\omega)=\frac{\Gamma_R}{2}F_R-{\rm i}\frac{\omega}{2}F_I$, with
\begin{eqnarray}\label{FRFI}
    F_R &=& \frac{\Gamma^2}{4}+\Delta^2 -\omega^2\pt{1+\frac{2\Gamma_d'}{\Gamma_R}+\frac{2T_c^2}{\Gamma_L}\frac{\Gamma_d'-\Gamma_L}{\Gamma_L^2+\omega^2}   } \,, \nonumber\\
    F_I&=&2T_c^2\Gamma_R\frac{\Gamma_L-\Gamma_d' }{\Gamma_L^2+\omega^2}
+\Gamma_d'\pt{\Gamma_d'+2\Gamma_R}
\nonumber\\&&
 +4T_c^2+\Delta^2-\omega^2\,.
\end{eqnarray}
In the following analysis we will also consider the situation, in which the resonator couples only with the left dot. In this case, the equations are modified accordingly, starting from the spectrum of the autocorrelation function of the electromechanical force, which in this latter case reads
\begin{equation}
S'(\omega)=-\alpha^2{\rm Tr}\left\{|L\rangle\langle L|\left({\cal L}_0\al{0}+{\rm i}\omega\right)^{-1}|L\rangle\langle L|\rho_{\rm St}\al{0}\right\}\,.
\end{equation}
The corresponding analytical form of the rates are not reported here, but are derived using the procedure described above.

\subsection{Cooling dynamics}

From the rate equation for the population of the phononic states one simply finds an equation for the mean number of mechanical excitations $\bar n=\sum_n np_n$, which reads~\cite{Eschner,Stenholm}
\begin{eqnarray}\label{nt}
\dot{\bar n}=-\gamma_{\rm tot}\bar n+A_+\,,
\end{eqnarray}
where $\gamma_{\rm tot}=A_--A_+$ is the cooling rate, and in particular
\begin{equation}\label{gammatot}
 \gamma_{\rm tot}=\gamma_p+\gamma_0
\end{equation}
with $\gamma_0=A_{0-}-A_{0+}$ the electromechanical cooling rate. A steady-state solution is found when $\gamma_{\rm tot}>0$, giving the average excitation number at steady state,
\begin{eqnarray}\label{n}
\bar n_{\rm St}= \frac{\bar n_p\gamma_p+\bar n_0\gamma_0 }{\gamma_p+\gamma_0}\,,
\end{eqnarray}
where $\bar n_0=A_{0+}/\gamma_0$ is the mean phononic number when $\gamma_p$ is set to zero.
The cooling dynamics results from the competition between the mechanical effect of the electron current and the thermalization with the environment. Clearly, the electromechanical coupling cools the resonator below the temperature $T$ of the environment when $\bar n_{\rm St}<\bar n_p$, which is possible when $\bar n_0<\bar n_p$. Equation~(\ref{n}) shows that lowest temperatures are achieved when $\gamma_0\gg\gamma_p$, giving
$$\bar n_{\rm St}\simeq \bar n_0+\pt{\bar n_p-\bar n_0}\gamma_p/\gamma_0\,.$$

In the following we will search for the parameters which minimize $\bar n_{\rm St}$ for different values of the thermalization rate $\gamma_p$ and of the dephasing rate $\Gamma_d$. In order to maximize the cooling rate, we first assume that $\Gamma_L$, the rate at which electrons are injected into the DQD from the left electrode, is the largest rate characterizing the dynamics (the higher bound is provided by the condition that the system must operate in the SET regime), implying that the time intervals, in which there is no excess electron inside the DQD, are here the smallest time scales. In the limit of large injection rates, the cooling rate $\gamma_0$ reads
\begin{widetext}
\begin{eqnarray} \label{W0}
&&\gamma_0\simeq\frac{\alpha^2}{\omega^2}\left(\frac{4T_c^2\Gamma_R}{\Delta^2
+ 2 T_c^2 + \frac{\Gamma_R^2}{4}}\right) \frac{\omega^3\Delta
\pt{2T_c^2+\Gamma_R^2+\omega^2}}
{\omega^2\pt{\Delta^2+4T_c^2+\frac{5}{4}\Gamma_R^2-\omega^2}^2+\Gamma_R^2\pt{\Delta^2+2T_c^2+\frac{\Gamma_R^2}{4}-2\omega^2}^2}+O\pt{\frac{1}{\Gamma_L}}\,,
\end{eqnarray} \end{widetext}
while the mean phonon number $\bar n_0$ takes the form
\begin{equation}
\bar n_{0}\simeq\frac{\Gamma_R^2+4\pt{\Delta-\omega}^2}{16\Delta\omega}+O\pt{\frac{1}{\Gamma_L}}\label{n0}\,.
\end{equation}
These expressions have been obtained setting the dephasing rate $\Gamma_d\ll\Gamma_R$. The effect of dephasing rates $\Gamma_d\simeq \Gamma_R$ on the cooling efficiency will be discussed later on. These expressions are valid both for the case in which the resonator couples to the right or to the left dot. Differences in the cooling efficiencies for the two setups arise when the value of $\Gamma_L$ is lowered, so that it becomes comparable with other rates, as we will show.

{\it Ground-state cooling.} Ground-state cooling may be obtained when $\bar n_0\ll 1$ and $\gamma_0\gg \gamma_p$. Assuming large injection rates into the DQD, and setting the dephasing rate $\Gamma_d=0$, small values $\bar n_0\ll 1$ are achieved in Eq.~(\ref{n0}) for $\omega\gg\Gamma_R$. The minimum value $$\bar
n_0=\Gamma_R^2/16\omega^2$$ is found setting $\Delta=\omega$. From Eq.~(\ref{W0}) we find that the cooling rate is maximum when tunneling rate and detuning fulfill the relation $\omega=\epsilon$ with
$$\epsilon\equiv\sqrt{\Delta^2+4T_c^2}\,.$$
This condition corresponds to the one which minimizes the temperature, provided that $|T_c|\ll\omega$. At sufficienctly large tunneling rates this condition emerges from the physical situation, in which the electronic states of the DQD are described by the bonding and antibonding states, such that $\epsilon$ is their frequency splitting. These states are the eigenstates of the Hamiltonian~(\ref{HDQD}), and are given by the quantum superposition of left and right electronic states
\begin{eqnarray}
    \ke{-}&=&\cos\theta\ke{L}+\sin\theta\ke{R}\nonumber\\ \ke{+}&=&-\sin\theta\ke{L}+\sin\theta\ke{R}
\end{eqnarray}
with $\tan\theta=2T_c/\pt{\Delta+\epsilon}$. In this basis there are two relevant physical processes, which lead to a change by one phonon due
to electron transport through the DQD. They consist of the sequential occupation of the states $|0, n\rangle \to |\pm, n\rangle \to |\mp,n\mp 1\rangle\to |0,n\mp 1\rangle$.
Both processes are resonant when the condition $\epsilon = \omega$ is satisfied. The resonator is cooled when the rate of the cooling process is faster than the heating
process, which is here satisfied for $\Delta > 0$, see Eq.~(\ref{W0}) and Ref.~\cite{Zippilli09}.

\begin{figure}[!th]
\begin{center}
\includegraphics[width=9cm]{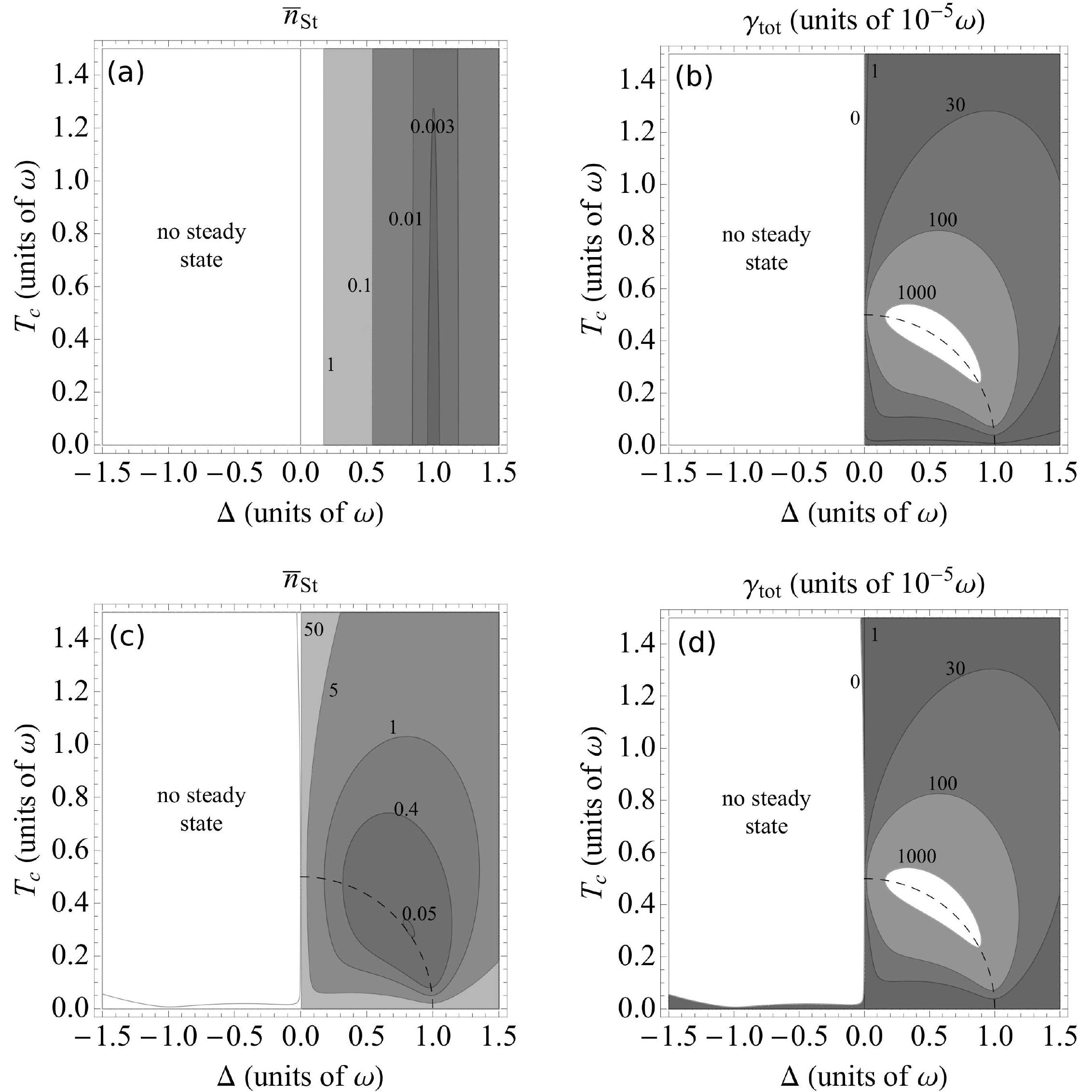}
\caption{Contour plots of the average phonon number at steady state $\bar n_{St}$, Eq.~(\ref{n}), and of the cooling rate $\gamma_{tot}$, Eq.~(\ref{nt}) as a function of $\Delta$ and $T_c$, in absence of dephasing. The grey scala is such, that the darkest (brightest) regions indicate the smallest (largest) values, with the exception of the region labeled with "no steady state", where the resonator is heated by electromechanical processes. The dashed line in the plots (b) and (d) indicates the curve $\Delta^2+4T_c^2=\omega^2$. The parameters are $\Gamma_L=100\omega$, $\Gamma_R=0.2\omega$, $\alpha =0.1\omega$, $\Gamma_d=0$ and (a)-(b) $\gamma_p=0$, (c)-(d) $\gamma_p=10^{-5}\omega$ with $\bar n_p=50$.
}\label{Fig:2}
\end{center}
\end{figure}

Figure~\ref{Fig:2} displays the contour plots for $\bar n_{\rm St}$ and $\gamma_{\rm tot}$ for a mechanical resonator with quality factor $Q=10^5$, the results are compared with the ideal situation $\gamma_p=0$ ($Q\to\infty$). We first discuss the ideal case. Here, we notice that the final mean occupation depends solely on the detuning $\Delta$, as expected from Eq.~(\ref{n0}), while large cooling rates are achieved when $T_c$ and $\Delta$ satisfy the resonance condition $\omega^2\simeq\Delta^2+4T_c^2$ and $0<\Delta<\omega$. At finite $Q$, efficient cooling is found for the condition at which the cooling rate  is maximized, as expected from Eq.~(\ref{n}). Note that the cooling efficiency decreases both in the limit $T_c\to0$ ($\Delta\to\omega$)  and $\Delta\to0$. In the first case, the total electron transport rate decreases, and correspondingly also the rate of electromechanical processes. In the second case, the asymmetry between heating and cooling rates ($A_\pm$) is reduced.

\begin{figure}[!th]
\begin{center}
\includegraphics[width=9cm]{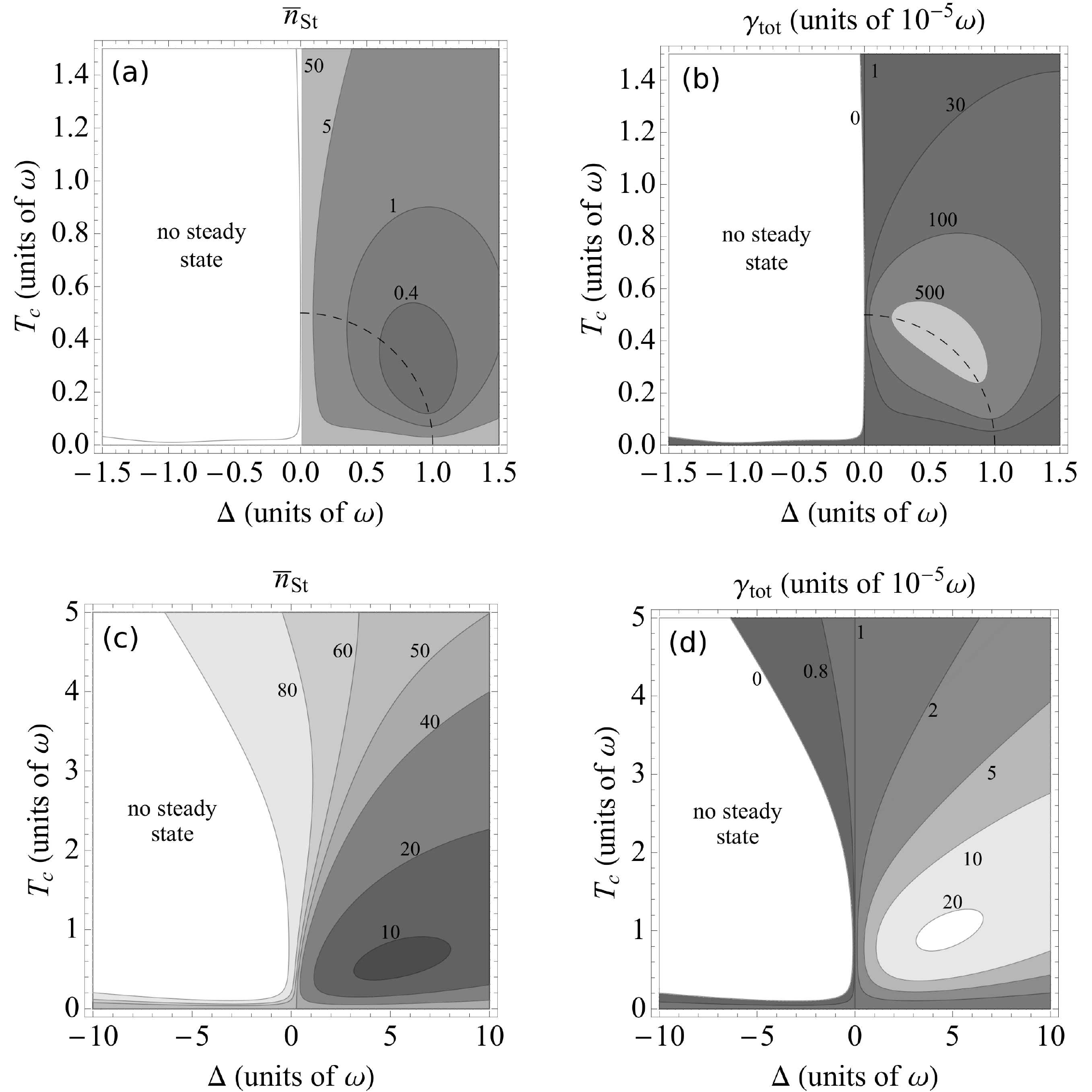}
\caption{Same as in Fig.~\ref{Fig:2}, but taking $Q =10^5$ ($\gamma_p = 10^{-5}\omega$) and $\bar n_p = 50$ everywhere and finite dephasing rate: In (a)-(b) $\Gamma_d=0.1\omega$, in (c)-(d) $\Gamma_d =5\omega$.
}\label{Fig:3}
\end{center}
\end{figure}

{\it Effect of dephasing on the cooling efficiency.} We now proceed in studying the effect of dephasing on the efficiency of the cooling mechanism. In Fig.~\ref{Fig:3} average phonon number and cooling rate are reported for $Q=10^5$ and setting $\Gamma_d=0.1\omega,5\omega$. Clearly, the cooling efficiency is lowered as the dephasing is increased.  In particular, the larger is the dephasing, the slower becomes the cooling. Nevertheless, for $\Gamma_d=5\omega$ it is found that the resonator can be cooled from $n_p=50$ to $n=10$ with a rate $\gamma_{\rm tot}=10^{-4}\omega$. We also observe that the parameter region, where cooling is achieved, is reduced (recall that the initial phonon number here considered is $n_p=50$). Optimal cooling is obtained for values of the detuning $\Delta$ of the order of $\Gamma_d$. Indeed, a simple calculation of the average phonon number, taking $\Gamma_d>\omega,\Gamma_R,T_c$, shows that this is modified according to the formula
\begin{eqnarray}
\bar n_{0}'\simeq \frac{\Gamma_d^2+\Delta^2}{2\Delta\omega}\,,
\end{eqnarray}
which reaches the minimal value $\bar n_0'\simeq \Gamma_d/\omega$ at $\Delta=\Gamma_d$. Correspondingly, the cooling rate is larger when $\omega>\Gamma_R$ and scales with $\gamma_0\sim \alpha^2 T_c^2/(\Gamma_d^2\omega)$. The cooling rate hence increases with the ratio between coherent tunneling and dephasing rate, with the upper bound $\gamma_0<\alpha^2/\omega$. This cooling limit is reminiscent of Doppler cooling of atoms in a dipolar transition with effective linewidth $2\Gamma_d$ and recoil frequency $\omega_R\sim \alpha^2/\omega$, see Refs.~\cite{Stenholm,Eschner}.

\begin{figure}[!th]
\begin{center}
\includegraphics[width=8.5cm]{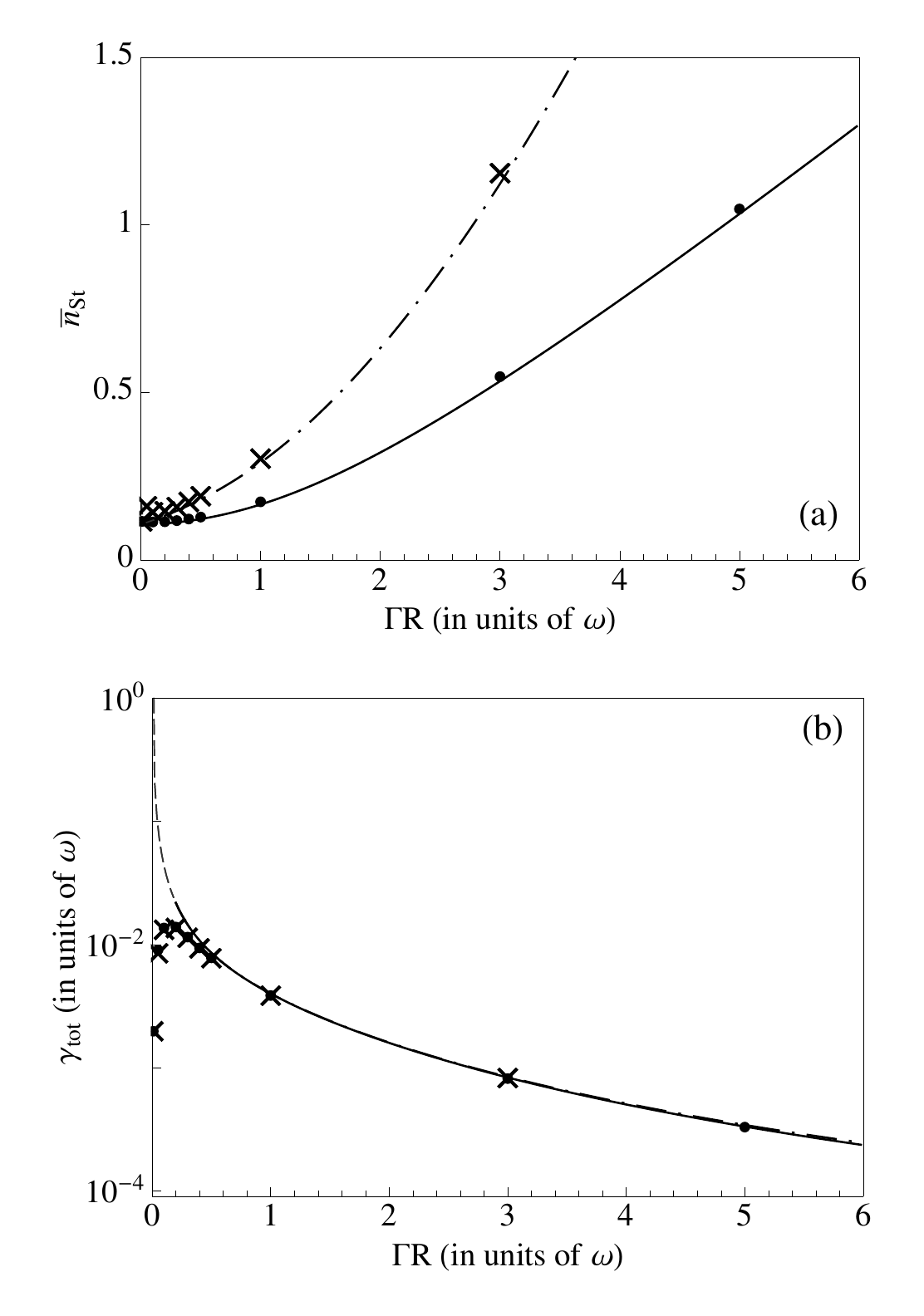}
\caption{(a) Average excitation number $\bar n_{\rm St}$ and (b) cooling rate $\gamma_{\rm tot}$ as a function of $\Gamma_R$ for $\gamma_p=0$. The solid (dash-dotted) curves are found for $\gamma_p=0$ ($\gamma_p=10^{-5}\omega$) by maximizing the values of $ \gamma_0$ as a function of $\Delta$ and $T_c$ in Eqs.~(\ref{n}) and~(\ref{gammatot}). The other parameters are $\alpha=0.1\omega$, $\Gamma_L=100\omega$, $\Gamma_d=0$ and $\bar n_p=50$. The dots and the crosses correspond to $\gamma_p=0$ and $\gamma_p=10^{-5}\omega$, respectively, and are extracted from numerical simulation, where the evolution of the resonator mean phonon number is calculated by numerical integration of master equation~(\ref{Meq}), for the same parameters as in the analytical case. The cooling rate is then determined by fitting the curve of the numerical integration with an exponential decay, minimizing the sum of the squares of the offsets of the numerical points from the exponential fit. The value of $\bar n_{\rm St}$ is extracted from the behaviour of the curve at sufficiently long evolution times. The dashed line in (b), visible at small $\Gamma_R$, indicates the regime in which the analytical results are not valid.  }
\label{Fig:4}
\end{center}
\end{figure}

{\it Cooling efficiency as a function of the incoherent tunneling rates.} So far we have assumed that injection rate into the DQD, $\Gamma_L$,  is the largest parameter, while we took $\Gamma_R<\omega$. We now study the cooling efficiency as a function of $\Gamma_L$ and $\Gamma_R$. We first focus on the situation in which $\Gamma_L$ is the largest parameter and consider the cooling rate as a function of $\Gamma_R$, when $\Gamma_d'\simeq \Gamma_R/2$. Figure~\ref{Fig:4} displays the average excitation number $n_{\rm St}$ and the cooling rate $\gamma_{\rm tot}$ as a function of $\Gamma_R$, choosing $T_c$ and $\Delta$ so to optimize both parameters. One clearly observes that optimal cooling is obtained at small values of $\Gamma_R$, while the cooling rate vanishes at $\Gamma_R\to0$. The numerical simulations show that the cooling efficiency has a maximum at $\Gamma_R$ of the order of a fraction of the oscillator frequency $\omega$, while it decreases as $\Gamma_R$ is further increased. We note, however, that phononic occupations $\bar n_{\rm St}<1$ are still obtained for $\Gamma_R\simeq \omega$. The analytical results reproduce well the behaviour at larger values of $\Gamma_R$, where one finds $\gamma_0\sim \Gamma_R^{-2}$. A significant discrepancy between analytical and numerical results is observed in the limit $\Gamma_R\to 0$. Such discrepancy is well understood, as in this regime the analytical treatment is invalid (as it is based on the approximation, that the dynamics of tunneling exceeds the rate of electromechanical coupling).

We have made further studies, considering the situation when the value of $\Gamma_L$ is such that the corrections scaling with $1/\Gamma_L$ are not negligible, for instance, choosing $\Gamma_L=10\omega$. In this case, we observe a significant decrease of the cooling efficiency. In particular, at lower values of $Q$ the scheme appears slightly more efficient when the resonator is coupled to the left dot rather than to the right dot.

\begin{figure}[!th]
\begin{center}
\includegraphics[width=8.5cm]{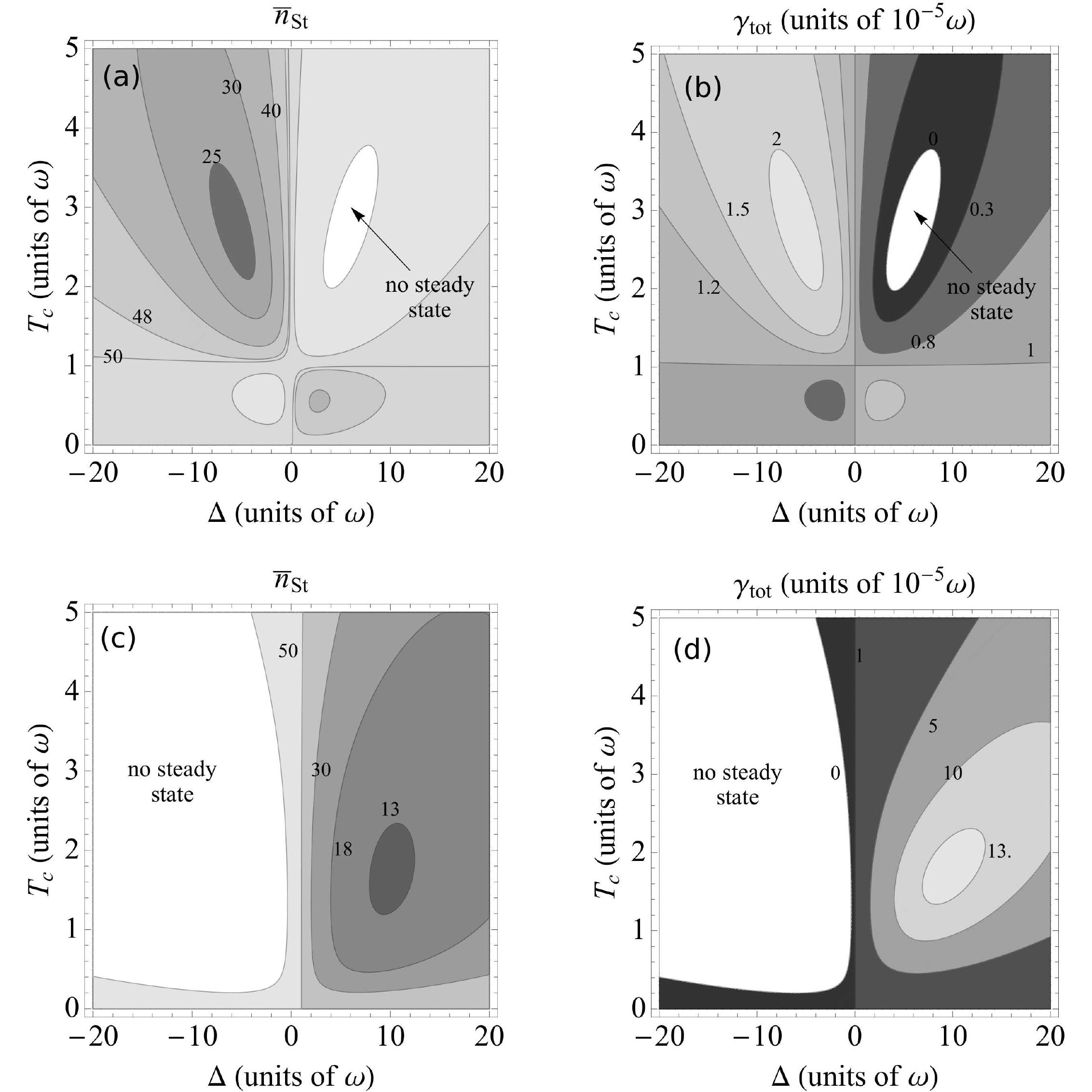}
\caption{Same as in Fig.~\ref{Fig:2}, but when the injection rate into the DQD is small, $\Gamma_L=0.1\omega$ and $\Gamma_R=10\omega$, such that DQD is essntially empty.
The other parameters are $\Gamma_R=10\omega$, $\alpha =0.1\omega$, $\Gamma_d=0.1\omega$, $\gamma_p=10^{-5}\omega$ and $\bar n_p=50$. In (a),(b) the resonator is coupled to the right dot, whereas in (c),(d) the resonator is coupled to the left dot.
}
\label{Fig:5}
\end{center}
\end{figure}

The dynamics is significantly modified, if one chooses $\Gamma_L$ small and $\Gamma_R$ comparatively large, see Fig.~\ref{Fig:5}. In this regime there is essentially no excess electron in the DQD at steady state. While $\bar n_0$ can be very small, the cooling rate is generally very slow, so that one obtains relatively large mean phononic numbers at finite values of $Q$. In this case, a slightly better cooling efficiency is found when the resonator is coupled to the left dot, see Fig.~\ref{Fig:5} (c) and (d).
We also note an unexpected behaviour: When the resonator is coupled to the right dot, cooling is found for negative values of $\Delta$, as visible in Figs.~\ref{Fig:5} (a) and (b) (hence, when the left dot is at higher energy than the right dot). This could be explained in terms of interference effects in the electromechanical coupling inside the DQD, analogously to the dynamics observed in Ref.~\cite{Morigi03,Zippilli05}.

\subsection{Discussion}
\label{App:Shift}

The scheme here proposed shares several analogies with laser cooling schemes of trapped ions, where the mechanical effects of photon-atom interaction is used in order to prepare the motion of trapped particles in the ground state of the confining potential~\cite{Stenholm,Eschner}. It is interesting to draw more explicit analogies, in order to evidentiate also the differences which arise when electromechanical effects are used in place of optomechanics.

A clear analogy between the two systems is found when considering the mechanical forces exerted on the harmonic oscillator. Indeed, the interaction Hamiltonian in Eq.~(\ref{He-ph}) can be mapped to the Hamiltonian, describing the mechanical effects by photon scattering in the Lamb-Dicke regime, by means of the unitary transformation~\cite{Wilson-Rae}
\begin{eqnarray}\label{Shift}
    S={\rm e}^{{\rm i}\frac{\alpha}{\omega}\kb{R}{R}\hat p}\,,
\end{eqnarray}
with $\hat p={\rm i}\pt{a\da-a}$
such that $\tilde\rho=S\da \rho S$, and the system operators transform as
\begin{eqnarray}
    S\da\kb{0}{R}S&=&\kb{0}{R}{\rm e}^{\frac{-{\rm i}\alpha}{\omega}\hat p}\nonumber\\
    S\da aS&=&a-\frac{\alpha}{\omega}\kb{R}{R}.
\end{eqnarray}
In this reference frame the interaction Hamiltonian is given by
\begin{eqnarray}
H_{Te-ph}=-T_c\pt{\kb{L}{R}{\rm e}^{-{\rm i}\frac{\alpha}{\omega}\hat p}+\kb{R}{L}{\rm e}^{{\rm i}\frac{\alpha}{\omega}\hat p} }
\end{eqnarray}
which has the same form as the interaction between a mode of the electromagnetic field and an atomic dipolar transition. In particular, in the limit of large $\Gamma_L$ the state $\ke{0}$, with no excess electrons in the DQD, can be adiabatically eliminated, and the DQD dynamics can be effectively described in the Hilbert space spanned by the state $\ke{L}$ and $\ke{R}$. In this regime, the system is analog to a trapped two-level atom with linewidth $\Gamma_d'$, cooled by a laser with Rabi frequency $T_c$ and possessing a Lamb-Dicke parameter $\alpha/\omega$~\cite{Stenholm,Eschner}. The dephasing rate hence plays a similar role of the linewidth of the cooling transition, setting the fundamental bound to the final temperature one can achieve by means of electromechanical forces in this setup. Drawing on this analogy, ground state cooling, as discussed in this paper, is a form of sideband cooling for trapped ions, and it is possible provided that the dephasing rate $\Gamma_d<\omega$. In that case, in this paper we show how the parameters can be tuned in order to achieve the largest ground state occupation. Summarizing, efficient cooling is obtained for large values of $\Gamma_L$ and small values of $\Gamma_R$, such that $\Gamma_R<\omega$. Large ground state occupations are also found when $\Gamma_R\simeq \omega$. When $\Gamma_d>\omega$, it is generally not possible to reach large ground-state occupations, even though cooling can be performed which has analogous efficiency as Doppler cooling.

Differing from photon scattering, in the case of electron tunneling we do not have diffusion processes, which are otherwise encountered when energy is dissipated by spontaneous emission in free space. For this reason,  Eq.~(\ref{n0}) has the same form found in sideband cooling of trapped ions under specific conditions, where diffusion due to photon scattering is suppressed~\cite{Cirac92,Zippilli05}. The same equation (for other physical parameters) was derived in theoretical treatments of cooling of nanomechanical resonators via photons~\cite{Kippenberg2,Marquardt,Genes}, where the cooling dynamics was mapped to sideband cooling of trapped ions. In our case, however, the parameters for which $\bar n_0$ is minimum do not coincide with the ones, at which $\gamma_0$ is maximum, and optimal cooling is found as a compromise between maximizing the electromechanical cooling rate, $\gamma_0$, and minimizing the lower bound to the mean phonon value, $\bar n_0$, so to effectively counteract the thermalization rate due to the coupling with the external environment.

\section{Conclusions}
\label{Sec:4}

We have presented an extensive analysis of the efficiency of ground state cooling of the phononic mode of a resonator, constituted by a suspended carbon-nanotube in a double quantum dot configuration. The analysis takes into account dephasing in the current through the double dot and the finite Q of the resonator. Ground state cooling is possible provided that the dephasing rate is smaller than the resonator frequency. Moreover, largest efficiency are obtained when the permanence time scale of the excess electron inside the double dot exceeds the time scale in which the double dot has no excess electron. When the dephasing rate is larger than the resonator frequency, the resonator can be still cooled provided that the tunable system parameters are accordingly chosen.

Most parameters of the proposed setup can be tuned. The frequency $\omega$ of the resonator, for instance, depends on the length of the nanotube section that is suspended. The frequency difference $\Delta$ and the tunneling rates $Tc$, $\Gamma_L$, and $\Gamma_R$, can be controlled by the external gates~\cite{Gotz}. Electron dephasing rate, however, remains to be measured in nanotubes. We remark that the proposed cooling scheme can be applied to other device layouts, such as mechanical resonators electrostatically coupled to fixed DQDs~\cite{Naik}.

\acknowledgements
We are grateful to Cecilia Lopez for discussions. We acknowledge support by the European Commission (EURYI; EMALI MRTN-CT-2006-035369; FP6-IST-021285-2), by the ESF (EUROQUAM, CMMC), and by the Spanish Ministerio de Ciencia y Innovaci\'on (QOIT, Consolider-Ingenio 2010; QNLP, FIS2007-66944; Ramon-y-Cajal; Juan-de-la-Cierva). G. M. acknowledges the German Research Council (DFG) for support (Heisenberg-professorship programme).

\end{document}